\documentclass[twocolumn,showpacs,prl,amsmath,amssymb]{revtex4}

\usepackage{graphicx}% Include figure files
\usepackage{dcolumn}% Align table columns on decimal point
\usepackage{bm}% bold math
\usepackage{amsmath,epsfig}

\newcommand{\ket}[1]{\left\vert#1\right\rangle}
\newcommand{\bra}[1]{\left\langle#1\right\vert}
\newcommand{\braket}[2]{\left\langle#1\vert#2\right\rangle}

\begin{document}

%\preprint{APS/123-QED}

\title{Irreversible decay of nonlocal entanglement via a "reservoir" of a single degree of freedom}
\author{A. Vaglica}%
\email{vaglica@fisica.unipa.it}
\author{G. Vetri}%
\affiliation{%
%Istituto Nazionale di Fisica della Materia and \\
Dipartimento di Scienze Fisiche ed Astronomiche and CNISM, Universit\`{a} di Palermo\\
via Archirafi 36, 90123 Palermo, Italy
}%

\date{\today}% It is always \today, today,
             %  but any date may be explicitly specified

\begin{abstract}
Recently, it has been realized that nonlocal disentanglement may
take a finite time as opposite to the asymptotic decay of local
coherences. We find in this paper that a sudden irreversible death
of entanglement takes place in a two atom optical Stern-Gerlach
model. In particular, the one degree non dissipative environment
here considered suddenly destroys the initial entanglement of any
Bell's states $\ket{\phi^{\pm}}$ superposition.
\end{abstract}

\maketitle
\section{\label{sec:level1} I. Introduction}
Superposition principle is a basic feature of Quantum Mechanics
(QM). When applied to composite systems superposition leads to
nonlocal coherences and quantum correlations (entanglement) which
have been extensively studied in the last years, and are even
expected of central interest for physics to come \cite{4h}.

As it is known, quantum coherences and entanglement are potential
resources in various quantum information processes, even if they
suffer any sort of environmental action which represents a serious
obstacle towards these applications. On the other hand, it is just
this fragility with respect to the environment, that is, the
propensity to disperse information throughout inaccessible degrees
of freedom that explicates the quantum-classical transition
\cite{z}.

An interesting point raised in \cite{di,eber1,do,da1,carv}
concerns the possible finite time disentanglement for bipartite
systems, as opposite to the usual local decoherence asymptotic
time. An experimental evidence of this peculiar trait of
entanglement has been reported recently by Almeida et al.
\cite{da2}. This issue has been analyzed \cite{eber1} in a simple
and realistic model where two initially entangled two-level atoms
separately interact with the multimode vacuum noise of two
distinct cavities. The authors find out that the nonlocal
decoherence may take place suddenly or at least as fast as the sum
of the normal single atom decay rates. This sudden death of
entanglement has been analyzed also for two Jaynes-Cummings (JC)
atoms \cite{eber2}. Also in this case the dynamics of the
entanglement between the atomic internal variables shows different
peculiarities for different initial states, with possible sudden
decays that are however followed by periodic revivals, due to
recovery of information by the system, being the cavities
lossless.

In this paper we wish to investigate if the separate non
dissipative interaction of each two-level atom with one only
degree of freedom but of continuous variables, may play the role
of the interaction with a reservoir leading to an irreversible
disentanglement of the bipartite system.

As well known, the optical Stern-Gerlach (OSG) model \cite{sle}
gives the opportunity of modifying the JC model by including the
interaction between the internal and the external atomic dynamics
via the e.m. mode of the cavity field, so actually providing the
coupling of each qubit to one only degree of freedom. It has
already been shown that this non dissipative coupling leads to a
decoherence in the dynamics of a single atom \cite{vag1,tvv1,ctvv}
and also affects the entanglement in the internal dynamics of two
atoms that successively interact with the same cavity \cite{tvv2}.
As will shown in what follows, the same interaction gives rise to
a disentanglement of the bipartite system which exhibits strong
analogies with the disentanglement generated by the vacuum noise
\cite{eber1,da2}.
\section{\label{sec:level2} II. Optical Stern-Gerlach model for two atoms}

Let us consider two identical isolated two-level atoms, $A$ and
$B$, respectively crossing two distinct ideal cavities, $a$ and
$b$ (in a direction orthogonal to the cavity axis), as in the OSG
model. These two uncoupled subsystems are described by the total
Hamiltonian
\begin{equation}\label{H}
\hat{H}=\hat{H}_{Aa}+\hat{H}_{Bb}
\end{equation}
where, in resonant conditions and in the rotating wave
approximation for both atoms,

\begin{subequations}\label{Hab}
\begin{eqnarray}
\hat{H}_{Aa}=\frac{\hat{p}_{A}^{2}}{2 m}+\hbar
\omega\left(\hat{a}^{\dag}\hat{a}+\hat{S}_{z}^{A}+\frac{1}{2}\right)\nonumber\\
+\hbar\varepsilon\sin(k\hat{x}_{A})\left(\hat{a}^{\dag}\hat{S}_{-}^{A}+\hat{S}_{+}^{A}\hat{a}\right)
\label{Haa}\\
 \hat{H}_{Bb}=\frac{\hat{p}_{B}^{2}}{2 m}+\hbar
\omega\left(\hat{b}^{\dag}\hat{b}+\hat{S}_{z}^{B}+\frac{1}{2}\right)\nonumber\\
+\hbar\varepsilon\sin(k\hat{x}_{B})\left(\hat{b}^{\dag}\hat{S}_{-}^{B}+\hat{S}_{+}^{B}\hat{b}\right)\label{Hbb}.
\end{eqnarray}
\end{subequations}
To simplify the notation we will sometimes use the indices $A$ and
$B$ to indicate the two subsystems $Aa$ and $Bb$, respectively.
The cavity fields are, as usually, described by the bosonic
operators $\hat{a}$, $\hat{a}^{\dag}$ and $\hat{b}$,
$\hat{b}^{\dag}$, and the two cavity frequencies $\omega = ck$, as
well as the coupling constants $\varepsilon$ for the two
subsystems, are assumed to be the same. The $1/2$ spin operators
$\hat{S}_{\pm}^{j}$, $\hat{S}_{z}^{j}$ account for the internal
dynamics of $j$-th atom, while the conjugate variables
$\hat{x}_{j}$ and $\hat{p}_{j}$ describe the external motion of
the same atom along the axis direction of the corresponding cavity
($j=A,B$).

When the initial uncertainties $\Delta x_{j}$ of the atomic wave
packets are sufficiently small with respect to the wavelength
$\lambda = 2\pi/k$ of both modes $a$ and $b$, the Hamiltonian
(\ref{H}) assumes the form (see, for example, Ref. \cite{cvv} for
one atom case)
\begin{equation}\label{Hlin}
\hat{H}=\frac{\hat{p}_{A}^{2}}{2 m}+\frac{\hat{p}_{B}^{2}}{2
m}+\hbar
\omega(\hat{N}_{A}+\hat{N}_{B})+\hbar\hat{\Omega}_{x}^{A}\hat{\mu}_{x}^{A}
+\hbar\hat{\Omega}_{x}^{B}\hat{\mu}_{x}^{B}
\end{equation}
where
\begin{equation}\label{cost}
\hat{\mu}_{x}^{A}=\frac{\hat{a}^{\dag}\hat{S}_{-}^{A}+\hat{S}_{+}^{A}\hat{a}}{2\sqrt{\hat{N}_{A}}},
\qquad\qquad
\hat{N}_{A}=\hat{a}^{\dag}\hat{a}+\hat{S}_{z}^{A}+\frac{1}{2},
\end{equation}
are constant of motion. Similarly for the subsystem $B$.
$\hat{\Omega}_{x}^{j}\equiv 2 \varepsilon k \sqrt{\hat{N}_{j}}
\hat{x}_{j}$ is the (operatorial) Rabi frequency depending on the
atomic position (with respect to a nodal point of the related
cavity mode function) and on the excitation number of the $j$-th
subsystem. Using previous results \cite{cvv} for the one atom
case, it is easy to show that the evolution operator may be
written as
\begin{eqnarray}\label{U}
\hat{U}(t,0)=e^{-i\hat{H}t/\hbar}=\exp\left\{-\frac{2it}{\hbar}m
\hat{a}_{N}^{A}\hat{\mu}_{x}^{A}\hat{x}_{A}\right\}\nonumber\\
\times\exp\left\{-\frac{it}{2 m \hbar} \hat{p}_{A}^{2}\right\}
\exp\left\{\frac{i t^{2}}{\hbar}
\hat{a}_{N}^{A}\hat{\mu}_{x}^{A}\hat{p}_{A}\right\}\nonumber\\
\times\exp\left\{-\frac{2it}{\hbar}m
\hat{a}_{N}^{B}\hat{\mu}_{x}^{B}\hat{x}_{B}\right\}
\exp\left\{-\frac{it}{2 m \hbar} \hat{p}_{B}^{2}\right\}\nonumber\\
\times\exp\left\{\frac{i t^{2}}{\hbar}
\hat{a}_{N}^{B}\hat{\mu}_{x}^{B}\hat{p}_{B}\right\}
e^{-i\vartheta_{0}(t)(\hat{N}_{A}+\hat{N}_{B})}
\end{eqnarray}
where
\begin{equation}\label{acc}
\hat{a}_{N}^{j}=a_{0}\sqrt{\hat{N}_{j}},\,\,
a_{0}=\frac{\varepsilon\hbar k}{m},\,\,\vartheta_{0}(t)=\omega t+
m a_{0}^{2}t^{3}/(6\hbar)
\end{equation}
and it has been assumed that each atom enters the own cavity at
the same time.

\section{\label{sec:level3} III. Reduced density operator for the qubits}
In this section we will analyze for the two different
configurations considered in Ref.\cite{eber2}, the time evolution
of the initial entanglement between the two qubits under the
effects of the interaction of each qubit with its own cavity and,
consequently, of the coupling with its own translational motion.
We will see that, differently from the usual JC model, in this
case the elements of the reduced density matrices that describe
the atomic internal dynamics show damped oscillations as a
consequence of the OSG effect which causes a splitting of the
atomic packets. As we will see later, this damping affects the
entanglement decay whose rate is actually related to the phase
space distance of the scattered packets. The section ends with a
brief account of the results for the one atom OSG model, which are
useful to explicate relations and differences between the decay
rates of nonlocal entanglement and local coherences.
\subsection{A. Two atoms}
Let us first suppose that the configuration of the entire system
at time $t=0$ may be written as
\begin{equation}\label{initot}
\ket{\psi(0)}=\ket{\varphi_{A}(0)}\ket{\varphi_{B}(0)}\ket{\chi(0)}
\end{equation}
where $\ket{\varphi_{A}}$, $\ket{\varphi_{B}}$ describe the
translational dynamics along the cavity axes of atoms $A$ and $B$,
while
\begin{equation}\label{chi0}
\ket{\chi(0)}=[\cos{\gamma}\ket{+-}+\sin{\gamma}\ket{-+}]\ket{00},\,\
(0\leq\gamma\leq\frac{\pi}{2})
\end{equation}
shows that the two qubits are initially in a superposition of the
Bell's states usually denoted $\ket{\psi^{\pm}}$, and the two
cavities are in the vacuum state. The ket $\ket{+-}$ indicates
that the qubit $A$ is in the upper state and  the qubit $B$ is in
the lover state, and so on. Using the dressed states
\begin{equation}\label{dre}
\ket{\chi_{j}^{\pm}}=\frac{1}{\sqrt{2}}[\ket{+0}_{j}\pm\ket{-1}_{j}],
\qquad j= A,B
\end{equation}
state (\ref{chi0}) assumes the form
\begin{eqnarray}\label{chi0dre}
\ket{\chi(0)}=\frac{1}{\sqrt{2}}\{\cos{\gamma}\left[\ket{\chi_{A}^{+}}+
\ket{\chi_{A}^{-}}\right]\ket{-0}_{B}\nonumber\\
+\sin{\gamma}\left[\ket{\chi_{B}^{+}}+\ket{\chi_{B}^{-}}\right]\ket{-0}_{A}\}.
\end{eqnarray}
Applying the evolution operator (\ref{U}) to the initial state
(\ref{initot}) and using Eq. (\ref{chi0dre}), one obtains the
following expression for the state of the entire system at time
$t$
\begin{eqnarray}\label{psit}
\ket{\psi(t)}=\frac{\cos{\gamma}}{\sqrt{2}}e^{-i\vartheta_{0}(t)}\ket{\varphi_{B}(t)}
[\ket{\phi_{A}^{+}(t)}\ket{\chi_{A}^{+}}\nonumber\\
+\ket{\phi_{A}^{-}(t)}\ket{\chi_{A}^{-}}]\ket{-0}_{B}\nonumber\\
+\frac{\sin{\gamma}}{\sqrt{2}}e^{-i\vartheta_{0}(t)}\ket{\varphi_{A}(t)}
[\ket{\phi_{B}^{+}(t)}\ket{\chi_{B}^{+}}\nonumber\\
+\ket{\phi_{B}^{-}(t)}\ket{\chi_{B}^{-}}]\ket{-0}_{A},
\end{eqnarray}
where we have set
\begin{eqnarray}
\ket{\phi_{j}^{\pm}(t)}= \exp\left\{\mp\frac{it}{\hbar}m
a_{0}\hat{x}_{j}\right\} \exp\left\{-\frac{it}{2 m \hbar}
\hat{p}_{j}^{2}\right\}\nonumber\\
\times\exp\left\{\pm\frac{it^{2}}{2\hbar}
a_{0}\hat{p}_{j}\right\}\ket{\varphi_{j}(0)}\qquad\qquad\label{fit1}\\
\ket{\varphi_{j}(t)}=\exp\left\{-i\frac{\hat{p}_{j}^{2}t}{2 m
\hbar}\right\}\ket{\varphi_{j}(0)}\qquad\qquad\;\;\;\;\;\label{fit2}
\end{eqnarray}
and the following relations have been used
\begin{eqnarray}\label{eigen}
\hat{N}_{j}\ket{\chi_{j}^{\pm}}=\ket{\chi_{j}^{\pm}},\,\,\hat{N}_{j}\ket{-0}_{j}=0,\nonumber\\
\hat{\mu}_{x}^{j}\ket{\chi_{j}^{\pm}}=\pm\frac{1}{2}\ket{\chi_{j}^{\pm}},\,\,
\hat{\mu}_{x}^{j}\ket{-0}_{j}=0.
\end{eqnarray}
If the initial spatial distributions of the two atoms are gaussian
functions of minimum uncertainty
\begin{equation}\label{gauss}
\varphi_{j}(x_{j},0)=\left(\frac{1}{\sqrt{2\pi}\Delta
x_{0}}\right)^{1/2}\exp\left\{-\frac{(x_{j}-x_{j,0})^{2}}{4\Delta
x_{0}^{2}}\right\}
\end{equation}
centered in $x_{j,0}$ and with the same spread $\Delta x_{0}$, the
$x$-representation of (\ref{fit1}) gives
\begin{eqnarray}\label{fitx}
\phi_{j}^{\pm}(x_{j},t)=\left[\frac{\Delta
x_{0}}{\sqrt{2\pi}\beta(t)}\right]^{1/2}\exp\left\{\mp\frac{it}{\hbar}m
a_{0}x_{j}\right\}\nonumber\\
\times\exp\left\{-\frac{\left[x_{j}-x_{j}^{\pm}(t)\right]^{2}}{4\beta(t)}\right\}\qquad\qquad
\end{eqnarray}
where
\begin{equation}\label{beta}
x_{j}^{\pm}(t)=x_{j,0}\mp a_{0}t^{2}/2,\qquad \beta(t)=\Delta
x_{0}^{2}+i\hbar t/(2m).
\end{equation}
As Eq.(\ref{fit2}) shows, $\ket{\varphi_{j}(t)}$ describes the
translational free motion of the $j$-th atom, while the
$x$-representation of $\ket{\phi_{j}^{\pm}(t)}$ given by
Eq.(\ref{fitx}) accounts for the well known splitting of the
incoming wave packet in the OSG effect.\\
Since we are interested in the entanglement of the two qubits, we
will identify the internal atomic degrees as the system of
interest, while the cavities and the external dynamics will serve
as environment. As a consequence, the density operator
$\ket{\psi(t)}\bra{\psi(t)}$ must be traced on the cavity fields
and on the atomic translation variables. Taking into account the
normalization of kets (\ref{fit1}) and (\ref{fit2}), and using
relation (\ref{dre}) we obtain
\begin{eqnarray}\label{roAB}
\rho^{AB}(t)=Tr_{transl, field}(\ket{\psi(t)}\bra{\psi(t)})\qquad\qquad\qquad\nonumber\\
=a_{1}\ket{+ -}\bra{+-}+ a_{2}\ket{-+}\bra{-+}+a_{3}\ket{+-}\bra{-+}\nonumber\\
+ a_{3}^{\ast}\ket{-+}\bra{+-}
+a_{4}\ket{--}\bra{--}\qquad\qquad\qquad
\end{eqnarray}
where
\begin{eqnarray}\label{coeff}
a_{1}=\frac{1}{2}\cos^{2}{\gamma}\left[1+Re\braket{\phi_{A}^{-}(t)}{\phi_{A}^{+}(t)}
\right]\qquad\qquad\qquad\nonumber\\
a_{2}=\frac{1}{2}\sin^{2}{\gamma}\left[1+Re\braket{\phi_{B}^{-}(t)}{\phi_{B}^{+}(t)}
\right]\qquad\qquad\qquad\nonumber\\
a_{3}=\frac{1}{4}\cos{\gamma}\sin{\gamma}\left(\braket{\varphi_{A}(t)}{\phi_{A}^{+}(t)}+
\braket{\varphi_{A}(t)}{\phi_{A}^{-}(t)}\right)\nonumber\\
\times\left(\braket{\phi_{B}^{-}(t)}{\varphi_{B}(t)}+
\braket{\phi_{B}^{+}(t)}{\varphi_{B}(t)}\right)\nonumber\\
a_{4}=\frac{1}{2}-\frac{1}{2}\cos^{2}{\gamma}Re\braket{\phi_{A}^{-}(t)}{\phi_{A}^{+}(t)}
\qquad\qquad\qquad\,\,\,\nonumber\\
-\frac{1}{2}\sin^{2}{\gamma}Re\braket{\phi_{B}^{-}(t)}{\phi_{B}^{+}(t)}.\qquad\qquad\qquad
\end{eqnarray}
By using the Eq.s (\ref{fit2}) and (\ref{fitx}) it is possible to
show that the following scalar products hold \cite{cvv,tvv1}
\begin{subequations}\label{scalprod}
\begin{eqnarray}
\braket{\phi_{j}^{-}(t)}{\phi_{j}^{+}(t)}=\exp\left\{-i\Omega_{j,0}t\right\}e^{-\Gamma(t)}\qquad\qquad\qquad\label{scalprod1}\\
\braket{\varphi_{j}(t)}{\phi_{j}^{\pm}(t)}=\exp\left\{i\frac{ma_{0}^{2}t^{3}}{4\hbar}\mp
i
\frac{\Omega_{j,0}t}{2}\right\}e^{-\Gamma(t)/4}\,\,\,\,\,\label{scalprod2}
\end{eqnarray}
\end{subequations}
where
\begin{equation}\label{gamma}
\Gamma(t)=\frac{\delta x(t)^{2}}{8\Delta x_{0}^{2}}+\frac{\delta
p(t)^{2}}{8\Delta p_{0}^{2}},
\end{equation}
$\Delta p_{0}$ is the initial uncertainty of the momentum
distributions for both atoms, and
\begin{equation}\label{dxdpomega}
\delta x(t)=-a_{0}t^{2},\; \delta p(t)=-2 m a_{0}t,\;
\Omega_{j,0}=2 m a_{0}x_{j,0}/\hbar.
\end{equation}
In the following we will assume that $x_{A,0}$ = $x_{B,0}$
$\equiv$ $x_{0}$, which implies
\begin{equation}\label{omega}
\Omega_{A,0}=\Omega_{B,0}\equiv\Omega_{0}=\frac{2 m}{\hbar}
a_{0}x_{0}.
\end{equation}
By inserting this equation and Eq.s (\ref{scalprod}) in
Eq.(\ref{coeff}) we get
\begin{subequations}\label{coeff2}
\begin{eqnarray}
a_{1}=\frac{1}{2}\cos^{2}{\gamma}\left\{1+\cos(\Omega_{0}t)e^{-\Gamma(t)}
\right\}\label{coeff21}\\
a_{2}=\frac{1}{2}\sin^{2}{\gamma}\left\{1+\cos(\Omega_{0}t)e^{-\Gamma(t)}
\right\}\label{coeff22}\\
a_{3}=\sin{\gamma}\cos{\gamma}\cos^{2}(\Omega_{0}t/2)e^{-\Gamma(t)/2}
\label{coeff23}\\
a_{4}=\frac{1}{2}\left\{1-\cos(\Omega_{0}t)e^{-\Gamma(t)}\right\}.\qquad
\label{coeff24}
\end{eqnarray}
\end{subequations}
It is useful to emphasize that $\Gamma(t)$ is proportional to the
square of the adimensional distance in phase space between the
average positions of scattered packets in the OSG effect, as
defined in Ref. \cite{ctvv} for the one atom case. The same
$\Gamma(t)$ can be also interpreted as the \textit{visibility} of
the Rabi oscillations \cite{tvv1} in the ambit of the
complementarity relation as introduced by Englert \cite{e}.

Considering the ordered basis $\ket{++}$, $\ket{+-}$, $\ket{-+}$,
$\ket{--}$, the reduced density operator (\ref{roAB}) takes on the
form
\begin{equation}\label{roABmatrix1}
\rho^{AB}(t)= \left( \begin{array}{cccc}
0 & 0 & 0 & 0 \\
0 & a_{1} & a_{3}& 0 \\
0 & a_{3}^{\ast} & a_{2}& 0 \\
0 & 0 & 0& a_{4}
\end{array} \right)
\end{equation}
where $(a_{1}+a_{2}+a_{4}) = 1$.
$a_{3}$ describes the coherence between the two atoms which
assumes its maximum value at $t=0$ and, contrarily to what happens
in the JC case, goes irreversibly to zero because of the coupling
of each qubit with its own external dynamics.

Now we consider the case in which the two qubits are initially in
a superposition of the Bell's states denoted $\ket{\phi^{\pm}}$,
and the two cavities are in the vacuum state,
\begin{equation}\label{chi02}
\ket{\chi(0)}=[\cos{\gamma}\ket{++}+\sin{\gamma}\ket{--}]\ket{00}.
\end{equation}
With respect to the more convenient ordered basis $\ket{+-}$,
$\ket{++}$, $\ket{--}$, $\ket{-+}$, the reduced density matrix of
the two qubits assumes the form
\begin{equation}\label{roABmatrix2}
\rho^{AB}(t)= \left( \begin{array}{cccc}
b_{3} & 0 & 0 & 0 \\
0 & b_{2} & b_{1}& 0 \\
0 & b_{1}^{\ast} & b_{5}& 0 \\
0 & 0 & 0& b_{3}
\end{array} \right)
\end{equation}
where
\begin{eqnarray}\label{coeffb1}
b_{1}=\frac{1}{4}\sin{\gamma}\cos{\gamma}e^{-i\vartheta_{0}(t)}\qquad\qquad\qquad\qquad\qquad\qquad\,\,\,\nonumber\\
\times\left(\braket{\varphi_{A}(t)}{\phi_{A}^{-}(t)}+
\braket{\varphi_{A}(t)}{\phi_{A}^{+}(t)}\right)\qquad\nonumber\\
\times\left(\braket{\phi_{B}^{-}(t)}{\varphi_{B}(t)}+
\braket{\phi_{B}^{+}(t)}{\varphi_{B}(t)}\right)\qquad\nonumber\\
b_{2}=\frac{\cos^{2}{\gamma}}{4}[1+Re\braket{\phi_{A}^{-}(t)}{\phi_{A}^{+}(t)}
+Re\braket{\phi_{B}^{-}(t)}{\phi_{B}^{+}(t)}\nonumber\\
+Re\braket{\phi_{A}^{-}(t)}{\phi_{A}^{+}(t)}
Re\braket{\phi_{B}^{-}(t)}{\phi_{B}^{+}(t)}]\nonumber\\
b_{3}=\frac{\cos^{2}{\gamma}}{4}[1+Re\braket{\phi_{A}^{-}(t)}{\phi_{A}^{+}(t)}-
Re\braket{\phi_{B}^{-}(t)}{\phi_{B}^{+}(t)}\nonumber\\
-Re\braket{\phi_{A}^{-}(t)}{\phi_{A}^{+}(t)}
Re\braket{\phi_{B}^{-}(t)}{\phi_{B}^{+}(t)}]\nonumber\\
b_{4}=\frac{\cos^{2}{\gamma}}{4}[1-Re\braket{\phi_{A}^{-}(t)}{\phi_{A}^{+}(t)}+
Re\braket{\phi_{B}^{-}(t)}{\phi_{B}^{+}(t)}\nonumber\\
-Re\braket{\phi_{A}^{-}(t)}{\phi_{A}^{+}(t)}
Re\braket{\phi_{B}^{-}(t)}{\phi_{B}^{+}(t)}]\nonumber\\
b_{5}=\frac{\cos^{2}{\gamma}}{4}[1-Re\braket{\phi_{A}^{-}(t)}{\phi_{A}^{+}(t)}-
Re\braket{\phi_{B}^{-}(t)}{\phi_{B}^{+}(t)}\nonumber\\
+Re\braket{\phi_{A}^{-}(t)}{\phi_{A}^{+}(t)}
Re\braket{\phi_{B}^{-}(t)}{\phi_{B}^{+}(t)}]+\sin^{2}{\gamma}.\nonumber\\
\end{eqnarray}
Using the results (\ref{scalprod}) we easily get
\begin{subequations}\label{coeff3}
\begin{eqnarray}
b_{1}=\sin{\gamma}\cos{\gamma}\cos^{2}(\Omega_{0}t/2)e^{-\Gamma(t)/2}\qquad\qquad\nonumber\\
\times\exp\left\{-i\left[\omega t - m
a_{0}^{2}t^{3}/(3\hbar)\right]\right\}\qquad
\label{coeff31}\\
b_{2}=\frac{1}{4}\cos^{2}{\gamma}\left\{1+\cos(\Omega_{0}t)e^{-\Gamma(t)}
\right\}^{2}\qquad\qquad\label{coeff32}\\
b_{3}=\frac{1}{4}\cos^{2}{\gamma}\left\{1-\cos^{2}(\Omega_{0}t)e^{-2\Gamma(t)}
\right\}\qquad\qquad\label{coeff33}\\
b_{5}=\sin^{2}{\gamma}+\frac{1}{4}\cos^{2}{\gamma}\left\{1-\cos(\Omega_{0}t)e^{-\Gamma(t)}
\right\}^{2} \label{coeff35}
\end{eqnarray}
\end{subequations}
where we have used $b_{4} = b_{3}$ because of the condition
(\ref{omega}).
\subsection{B. One atom}
To analyze analogies and differences with the disentanglement
produced by different environments, it is useful to report in
short on the dynamics of a single two-level atom in the OSG model.
We assume that the initial state of whole system is
\begin{equation}\label{initot1}
\ket{\psi(0)}=\ket{\varphi(0)}\ket{\chi(0)}
\end{equation}
where
\begin{equation}\label{chi01}
\ket{\chi(0)}=[\cos{\gamma}\ket{+}+\sin{\gamma}\ket{-}]\ket{0}
\end{equation}
and $\ket{\varphi}$ describes the atomic translational dynamics
along the cavity axis. Using some results of Ref.\cite{cvv} (see
sec. IV) one obtains the time evolution of (\ref{initot1}),
\begin{eqnarray}\label{psit}
\ket{\psi(t)}=\sin{\gamma}\ket{\varphi(t)}\ket{-0}\qquad\qquad\qquad\qquad\qquad\nonumber\\
+\frac{\cos{\gamma}}{\sqrt{2}}e^{-i\vartheta_{0}(t)}
[\ket{\phi^{+}(t)}\ket{\chi^{+}}
+\ket{\phi^{-}(t)}\ket{\chi^{-}}].
\end{eqnarray}
Removing the environment by tracing on the field and on the atomic
translation variables, one gets the reduced one qubit density
matrix
\begin{equation}\label{romatrix1}
\rho(t)= \left( \begin{array}{cc}
q_{1}  & q_{2} \\
q_{2}^{\ast} & q_{3}
\end{array} \right)
\end{equation}
where
\begin{subequations}\label{coeffq}
\begin{eqnarray}
q_{1}=\frac{1}{2}\cos^{2}{\gamma}\left[1+Re\braket{\phi^{-}(t)}{\phi^{+}(t)}
\right]\qquad\qquad\qquad\qquad\label{coeffq1}\\
q_{2}=\frac{1}{2}e^{-i\vartheta_{0}(t)}\cos{\gamma}\sin{\gamma}\left[\braket{\varphi(t)}{\phi^{+}(t)}+
\braket{\varphi(t)}{\phi^{-}(t)}\right]\nonumber\\
\label{coeffq2}\\
q_{3}=\frac{1}{2}\cos^{2}{\gamma}\left[1-Re\braket{\phi^{-}(t)}{\phi^{+}(t)}
\right]+\sin^{2}{\gamma}.\qquad\qquad\label{coeffq3}
\end{eqnarray}
\end{subequations}
Both the irreversible evolution of the populations
((\ref{coeffq1}) and (\ref{coeffq3})) and the decoherence
(\ref{coeffq2}) are to be ascribed to the increasing distance in
the phase space between the scattered atomic packets. However,
while the population decay depends on the square distance
$\Gamma(t)$ between the packets moving in opposite directions
(through the scalar product $\braket{\phi^{\mp}}{\phi^{\pm}}$),
the coherence experiences a four time slower decay since it
depends on the square distance $\Gamma(t)/4$ between one of the
moving packet and the standing packet related to the ground state
(scalar product $\braket{\varphi}{\phi^{\pm}}$). It is moreover to
stress that, because of the non dissipative character of the
decay, the qubit upper population does not go to zero, as in the
spontaneous emission, but tends to half of the initial value.
\section{\label{sec:level4} IV. Entanglement evolution for the two qubits}
\subsection{A. Concurrence}
\begin{figure}
 \includegraphics[width = 0.42\textwidth]{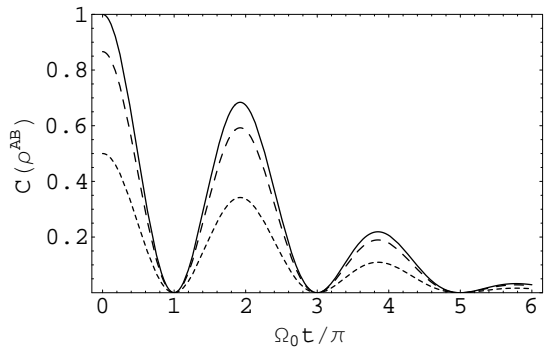}
 \caption{\label{fig1} Decay of concurrence, as given by Eq.
 (\ref{concu}),
when the two qubits are initially in the superposition
(\ref{chi0}) of Bell's states $\ket{\psi^{\pm}}$. The
superposition parameter is $\gamma=\pi/4$ (continuous line),
$\pi/6$ (dashed line), $\pi/12$ (dotted line), with
$x_{A,0}=x_{B,0}\equiv x_{0}= \lambda/10$, $\Delta
x_{0}=\lambda/50$. The other parameters are $\lambda=10^{-2} meter
$, $\epsilon=10^{4} sec^{-1}$, $m=10^{-26} kg$.}
\end{figure}

As it is generally convenient for these bipartite mixed states we
will analyze the time evolution of the entanglement between the
internal variables of the two atoms, looking at the concurrence
\cite{w}. It is not difficult to show that the concurrence of the
state (\ref{roABmatrix1}) is
\begin{eqnarray}
C(\rho^{AB})=max\{0,
\left(\left||a_{3}|+\sqrt{a_{1}a_{2}}\right|-\left||a_{3}|
-\sqrt{a_{1}a_{2}}\right|\right)\}\nonumber\\
=2a_{3}.\qquad\qquad\qquad\qquad\qquad\qquad\qquad\qquad\label{concu}
\end{eqnarray}
The entanglement with the non relevant systems causes a
degradation of the nonlocal correlation between the qubits.
According to previous results (see, for example \cite{eber1,da1}),
and as shown in Fig.\ref{fig1}, state (\ref{chi0}) gives rise to
asymptotic disentanglement with the same rate of the decoherence.
Because of the particular dynamics of the OSG model, in this case
the disentanglement rate is one half time the population decay
rate. Finally, we note that, for $\Gamma(t)=0$, i.e. when
translation dynamics of the center of mass of the two atoms is not
taken into account, the concurrence (\ref{concu}) reduces to that
derived in \cite{eber2} for the two JC atoms case.

\begin{figure}
 \includegraphics[width = .42\textwidth]{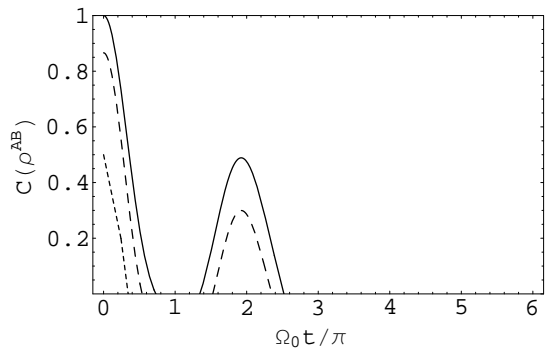}
 \caption{\label{fig2} Decay of concurrence, as given by Eq. (\ref{concu2}), when
the two qubits are initially in the superposition (\ref{chi02}) of
Bell's states $\ket{\phi^{\pm}}$. The values of the parameters are
as in figure 1.}
\end{figure}

More interesting results are expected  for the initial state
(\ref{chi02}), for which the concurrence is
\begin{equation}\label{concu2}
C(\rho^{AB})=max\left\{0,d(t)\right\}
\end{equation}
where
\begin{eqnarray}\label{concu2d}
d(t)=|b_{1}|+\sqrt{b_{2}b_{5}}-||b_{1}|-\sqrt{b_{2}b_{5}}|-2b_{3}\nonumber\\
=2(|b_{1}|-b_{3}).\qquad\qquad\qquad\qquad\qquad
\end{eqnarray}
The behavior of concurrence (\ref{concu2}) is shown in Fig.2. Also
in this case, the coherence $b_{1}$ between the two atoms goes to
zero as $a_{3}$ in the previous case does. However, the
concurrence experiences in this case a sudden death because of the
probabilities $b_{3}$ of finding the atoms in the two states
$\ket{+-}$ and $\ket{-+}$ which are not correlated to the others
and which tend to $\cos^{2}(\gamma)/4$ for long time. Figure
(\ref{fig2}) makes evident the strong dependence of the
entanglement life-time on the parameter of the initial two qubit
state (\ref{chi02}), similarly to the theoretical \cite{eber1} and
experimental \cite{da2} results. However, in this two last papers
it is outlined that only some initial entanglements may undergo to
sudden death. On the contrary, due to the particular environmental
action of the OSG model, there is no threshold in our case. In
fact, the function $ d(t)$ of Eq. (\ref{concu2d}) is negative if $
b_{3}/|b_{1}|>1$ and it turns out that
\begin{eqnarray}\label{dis}
b_{3}/|b_{1}|\geq\frac{\cot\gamma}{2}\sinh\left[\Gamma(t)/2\right].\qquad\qquad\qquad
\end{eqnarray}
From this equation one can realize that for any value of $\gamma$
there exists a finite time for which the concurrence goes to
negative values and the two qubits become disentangled. In this
respect, our model behaves similarly to the continuous variable
two-atom model of Ref. \cite{do}.
\subsection{B. Separability}
We will now inquire on the separability of the reduced density
matrices (\ref{roABmatrix1}) and (\ref{roABmatrix2}) adopting the
Peres-Horodecky test \cite{p,h}. As it is known, for two qubits
the non-negativity of the eigenvalues of the partial transposed
matrix works as a necessary and sufficient condition for the
separability. Consider first the partial transposed of
(\ref{roABmatrix1}),
\begin{equation}\label{sigABmatrix1}
\sigma^{AB}(t)= \left( \begin{array}{cccc}
0 & 0 & 0 & a_{3}^{\ast} \\
0 & a_{1} & 0 & 0 \\
0 & 0 & a_{2}& 0 \\
a_{3} & 0 & 0 & a_{4}
\end{array} \right)
\end{equation}
whose elements are given by Eq.s (\ref{coeff2}). We easily find
that three eigenvalues of (\ref{sigABmatrix1}),
\begin{figure}
 \includegraphics[width = .42\textwidth]{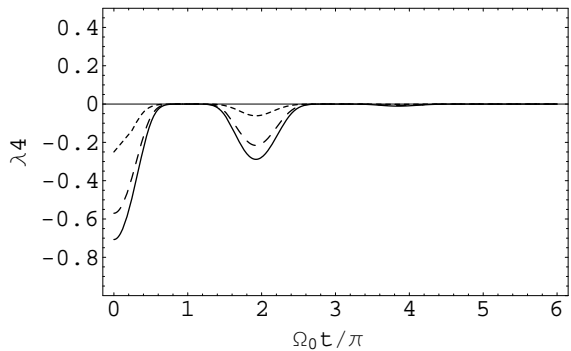}
 \caption{\label{fig3} Time behavior of the eigenvalue (\ref{eig12})
 of the matrix (\ref{sigABmatrix1}) for the one excitation initial state (\ref{chi0}).
 An enlarged form of this figure should show that $\lambda_{4}$ takes on the zero value only for isolated
 points. The values of the parameters are as in figure 1.}
\end{figure}
\begin{figure}
 \includegraphics[width = .42\textwidth]{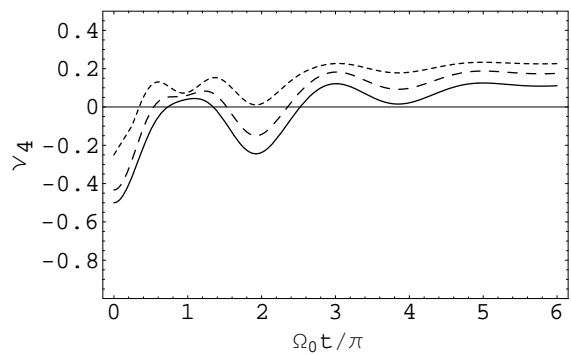}
 \caption{\label{fig4} Time behavior of the forth eigenvalue (\ref{eig22})
 of the matrix (\ref{sigABmatrix2}) for the two excitations initial state (\ref{chi02}).
 The values of the parameters are as in figure 1.}
\end{figure}
\begin{eqnarray}
\lambda_{1}=a_{1},\,\,\,\,\lambda_{2}=a_{2},\,\,\,\,
\lambda_{3}=\frac{a_{4}}{2}+\frac{1}{2}\left[a_{4}^{2}+4a_{3}^{2}\right]^{1/2},\label{eig11}
\end{eqnarray}
are non-negative, while the forth eigenvalue
\begin{equation}\label{eig12}
\lambda_{4}=\frac{a_{4}}{2}-\frac{1}{2}\left[a_{4}^{2}+4a_{3}^{2}\right]^{1/2}
\end{equation}
assumes only non-positive values, as figure (\ref{fig3}) shows.

We wish to point out that for the initial state (\ref{chi0}) the
two qubits display EPR correlations \cite{we} for all times, but
for isolated points (see Fig.(\ref{fig1})). Accordingly, one can
see from the analytical expression of $\lambda_{4}$ (more
accurately than from Fig. (\ref{fig3})) that the bipartite system
is not separable but for isolated points, and tends to the
separability for long times.

Finally, for the density matrix (\ref{roABmatrix2}) relative to
the initial state (\ref{chi02}) the partial transposition gives
\begin{equation}\label{sigABmatrix2}
\sigma^{AB}(t)= \left( \begin{array}{cccc}
b_{3} & 0 & 0 & b_{1}^{\ast} \\
0 & b_{2} & 0 & 0 \\
0 & 0 & b_{5}& 0 \\
b_{1} & 0 & 0 & b_{3}
\end{array} \right)
\end{equation}
whose elements are given in this case by Eq.s (\ref{coeff3}). As
in the previous case, three eigenvalues
\begin{equation}\label{eig21}
\nu_{1}=b_{2},\qquad\nu_{2}=b_{5},\quad\nu_{3}=b_{3}+|b_{1}|
\end{equation}
are non-negative, while a forth eigenvalue
\begin{equation}\label{eig22}
\nu_{4}=b_{3}-|b_{1}|
\end{equation}
takes on negative values for some time intervals (see figure
(\ref{fig4})), in exact correspondence with the existence of a
concurrence (see figure (\ref{fig2})).
\section{\label{sec:level5} V. Conclusions}
In this paper it has been analytically derived the time evolution
of the initial entanglement of two atoms which separately interact
with two distinct cavities. We have used the optical Stern-Gerlach
model to include the external translational dynamics of each atom.
The presence of entanglement has been ascertained trough the
concurrence and the separability and we have found, as in Ref.
\cite{eber1}, that the two atom internal dynamics follows
different patterns and becomes irreversibly separable in different
times, depending on the different initial configurations. As found
in other contests \cite{eber1,da1,da2} which use different
environments and where the disentanglement is due essentially to
spontaneous emission, also in the OSG model the entanglement may
undergo to an irreversible sudden death. In addition, the
particular non dissipative environment here considered suddenly
destroys the initial entanglement of Bell's states
$\ket{\phi^{\pm}}$ superposition (Eq. \ref{chi02}) for any initial
value of $\gamma$. This can in part be ascribed to the peculiar
dynamics of the qubits in the OSG model in which the population
$b_{3}$ (matrix (\ref{roABmatrix2})) of the non correlates states
$\ket{+-}$ and $\ket{-+}$ does not go to zero, but tend to
$\cos^{2}(\gamma)/4$ for long time. We can consequently conclude
that nonlocal coherences manifest an amazing fragility even
towards a non dissipative environment made of a single degree of
freedom.

\end{document}